\documentclass[graybox]{svmult}

\usepackage{mathptmx}       % selects Times Roman as basic font
\usepackage{helvet}         % selects Helvetica as sans-serif font
\usepackage{courier}        % selects Courier as typewriter font
\usepackage{type1cm}        % activate if the above 3 fonts are
\usepackage{makeidx}         % allows index generation
\usepackage{graphicx}        % standard LaTeX graphics tool
\usepackage{multicol}        % used for the two-column index
\usepackage[bottom]{footmisc}% places footnotes at page bottom

\usepackage{amsmath,amssymb}
\newcommand{\Real}{\mathbb{R}}

\newcommand{\pmu}{\partial_{\mu}}
\newcommand{\pnu}{\partial_{\nu}}
\newcommand{\nmu}{\nabla_{\mu}}
\newcommand{\nnu}{\nabla_{\nu}}
\newcommand{\FF}{\mathcal{F}}

\begin{document}

\title*{On Nonlocal Modified Gravity and Cosmology}
\author{Branko Dragovich   \\
    Institute of Physics, University of Belgrade,  Belgrade, Serbia}
%\institute{B. Dragovich \at Institute of Physics, University of Belgrade, Pregrevica 118, 11080 Zemun, Belgrade, Serbia
 %\email{dragovich@ipb.ac.rs}}
%\and Name of Second Author \at Name, Address of Institute \email{name@email.address}}
%
%
\maketitle

\vskip-3cm

\abstract{Despite many nice properties and numerous achievements, general relativity
is not a complete theory. One of actual approaches towards more complete theory of gravity
is its nonlocal modification. We present here a brief review of nonlocal gravity  with its
cosmological solutions. In particular, we pay special attention to two nonlocal models and their nonsingular bounce
solutions for the cosmic scale factor.}

\section{Introduction}
\label{sec:1}

Recall that General Relativity is the Einstein theory of gravity based on tensorial equation of motion for gravitational
(metric) field $g_{\mu\nu}: \quad R_{\mu\nu} - \frac{1}{2} R g_{\mu\nu} = {8 \pi G} T_{\mu\nu},$ where $R_{\mu\nu}$ is the
Ricci curvature tensor, $R$ -- the Ricci scalar, $T_{\mu\nu}$ is the energy-momentum tensor, and speed of light is $c = 1$. This Einstein equation
follows from the Einstein-Hilbert action $S = \frac{1}{16\pi G} \int \sqrt{-g}\, R \, d^4x + \int \sqrt{-g} \mathcal{L}_m \, d^4,$
where $g = \det(g_{\mu\nu})$ and $\mathcal{L}_m$ is Lagrangian of matter.

Motivations for modification of general relativity are usually related to some problems in quantum gravity, string theory, astrophysics and
cosmology (for a review, see \cite{clifton,nojiri, faraoni}).
We are here mainly interested in cosmological reasons to modify the Einstein theory of gravity. If general relativity is gravity theory for the universe
as a whole and the universe has Friedmann-Lema\^{\i}tre-Robertson-Walker (FLRW) metric, then there is in the universe about $68\%$ of {\it dark energy},
$27\%$ of {\it dark matter}, and only $5\%$ of {\it visible matter} \cite{planck}. The visible matter is described by the
Standard model of particle physics. However, existence of this $95\%$ of dark energy-matter content of the universe is still hypothetical, because it
has been not verified in the laboratory ambient. Another cosmological problem is related to the Big Bang singularity. Namely, under rather general
conditions, general relativity yields cosmological solutions with zero size of the universe at its beginning, what means an infinite matter density.
Note that when physical theory contains singularity, it is not valid in the vicinity of singularity and must be appropriately  modified.

In this article, we briefly review nonlocal modification of general relativity in a way to point out cosmological solutions without Big Bang
singularity. We consider two nonlocal models and present their nonsingular bounce cosmological solutions. To have more complete view of these models
we also write down  other
exact solutions which are power-law singular ones of the form $a(t) = a_0 \, |t|^\alpha .$

In Section 2 we describe some general characteristics of nonlocal gravity which are useful for understanding what follows in the sequel.
Section 3 contains a review of both nonsingular bounce and singular cosmological solutions for two nonlocal gravity models without matter.
Last section is related to the discussion with some concluding remarks.

\section{Nonlocal Gravity}
\label{sec:2}

 The well founded modification of the Einstein theory of gravity has to contain general relativity and to be verified on the dynamics of the Solar system.
 Mathematically, it should be formulated within the pseudo-Riemannian geometry in terms of covariant quantities and equivalence of the inertial and
 gravitational mass. Consequently, the Ricci scalar $R$ in gravity Lagrangian $\mathcal{L}_g$ of the Einstein-Hilbert action  has to be replaced  by a
 function which, in general, may contain not only $R$ but also any covariant construction which is possible in the Riemannian geometry. Unfortunately,
 there are infinitely many such possibilities  and so far without  a profound theoretical principle which could make definite choice. The Einstein-Hilbert
 action can be viewed as a result of the principle of simplicity  in construction of $\mathcal{L}_g$.

We consider here nonlocal modified gravity.
In general, a nonlocal modified gravity model corresponds to an infinite number of spacetime derivatives in the form  of some power expansions of the d'Alembert
operator $\Box = \frac{1}{\sqrt{-g}} \partial_{\mu}\sqrt{-g} g^{\mu\nu} \partial_{\nu}$ or of its inverse $\Box^{-1},$ or  some combination of both.
We are mainly interested in nonlocality expressed in the form of an analytic function $ \mathcal{F}(\Box)=  \sum_{n =0}^{\infty} f_{n}\Box^{n}.$  However,
some models with
$\Box^{-1} R,$ have been also considered (see, e.g. \cite{woodard,woodard-d,woodard1,nojiri1,nojiri2,sasaki,vernov0,vernov1,koivisto,koivisto1} and references therein). For nonlocal gravity with
$\Box^{-1}$ see also \cite{barvinsky,modesto}. Many aspects of nonlocal gravity models have been considered, see e.g.
\cite{modesto1,modesto2,moffat,calcagni,maggiore} and references therein.

Motivation to modify gravity in a nonlocal way comes mainly from string theory. Namely, strings are one-dimensional extended objects and  their field
theory description contains spacetime nonlocality. We will discuss it in the framework of $p$-adic string theory in Section 4.

In order to better understand  nonlocal modified gravity itself, we investigate it without matter. Models of nonlocal gravity which we mainly consider
are given by the action
\begin{equation}
S =  \int d^{4}x \sqrt{-g}\Big(\frac{R - 2 \Lambda}{16 \pi G} +  R^{q}
\mathcal{F}(\Box) R  \Big),   \quad q = +1, -1,          \label{eq:2.1}
\end{equation}
where $\Lambda$ is cosmological constant, which is for the first time introduced by Einstein in 1917. Thus this nonlocality is given by the term $R^{q}
\mathcal{F}(\Box) R , $ where $q= \pm 1$ and $ \mathcal{F}(\Box)=  \sum_{n =0}^{\infty} f_{n}\Box^{n},$ i.e. we investigate two nonlocal gravity
models: the first one with $q = + 1$ and  the second one with $q = - 1.$

Before to proceed, it is worth mentioning that analytic function $ \mathcal{F}(\Box)=  \sum_{n =0}^{\infty} f_{n}\Box^{n},$ has to satisfy some conditions,
in order to escape unphysical degrees of freedom  like ghosts and tachyons, and to be asymptotically free in the ultraviolet region
(see discussion in \cite{biswas3,biswas4}).

\section{Models and Their Cosmological Solutions}

In the sequel we shall consider the above mentioned two nonlocal models \eqref{eq:2.1} separately for $q = +1$ and $q = - 1 .$

We use the FLRW metric
$ds^2 = - dt^2 + a^2(t)\big(\frac{dr^2}{1-k r^2} + r^2 d\theta^2 +
r^2 \sin^2 \theta d\phi^2\big)$ and investigate all three
possibilities for curvature parameter $k =0,\pm 1$. In the FLRW
metric scalar curvature is $R = 6 \left (\frac{\ddot{a}}{a} +
\frac{\dot{a}^{2}}{a^{2}} + \frac{k}{a^{2}}\right )$ and $\Box =
- \partial_t^2  - 3 H \partial_t  ,$ where $H =
\frac{\dot{a}}{a}$ is the Hubble parameter. Note that we use natural system of units in which
speed of light $c = 1.$

\subsection{Nonlocal Model Quadratic in $R$}

Nonlocal gravity model which is quadratic in $R$ is given by the action \cite{biswas1,biswas2}
\begin{equation}
S =  \int d^{4}x \sqrt{-g}\Big(\frac{R - 2 \Lambda}{16 \pi G} +  R
\mathcal{F}(\Box) R  \Big).             \label{eq:3.1}
\end{equation}
This model is attractive because it is ghost free and has some nonsingular bounce solutions, which can solve
the Big Bang cosmological singularity problem.

The corresponding equation of motion follows from the variation of the action (\ref{eq:3.1}) with
respect to metric $g_{\mu\nu}$ and it is
\begin{eqnarray}
&  2 R_{\mu\nu} \mathcal{F}(\Box) R - 2(\nmu\nnu -
g_{\mu\nu} \Box)( \mathcal{F}(\Box) R) - \frac{1}{2}
g_{\mu\nu} R \mathcal{F}(\Box) R \nonumber \\
&+ \sum_{n=1}^{\infty} \frac{f_n}{2} \sum_{l=0}^{n-1} \big(
g_{\mu\nu} \left( g^{\alpha\beta}\partial_{\alpha} \Box^l R
\partial_{\beta} \Box^{n-1-l} R + \Box^l R \Box^{n-l} R
\right) \nonumber \\
&- 2 \pmu \Box^l R \pnu \Box^{n-1-l} R\big)  = \frac{-1}{8
\pi G} (G_{\mu\nu} + \Lambda g_{\mu\nu}). \label{eq:3.2}
\end{eqnarray}

When metric is of the FLRW form  in \eqref{eq:3.2} then there are only two
independent equations.
It is practical to use the trace  and $00$ component of \eqref{eq:3.2}, and respectively they  are:
\begin{eqnarray}
&6\Box ( \mathcal{F}(\Box) R) + \sum_{n=1}^{\infty} f_n
\sum_{l=0}^{n-1} \left(
\partial_{\mu} \Box^l R
\partial^{\mu} \Box^{n-1-l} R + 2 \Box^l R \Box^{n-l} R
\right)\nonumber \\ &= \frac{1}{8 \pi G} R - \frac{\Lambda}{2 \pi G}, \label{eq:3.3}
\end{eqnarray}
\begin{eqnarray}
&  2 R_{00} \mathcal{F}(\Box) R - 2(\nabla_0
\nabla_0 - g_{00} \Box)( \mathcal{F}(\Box) R) - \frac{1}{2}
g_{00} R \mathcal{F}(\Box) R  \nonumber \\
&+ \sum_{n=1}^{\infty} \frac{f_n}{2} \sum_{l=0}^{n-1} \big( g_{00}
\left( g^{\alpha\beta}\partial_{\alpha} \Box^l R
\partial_{\beta} \Box^{n-1-l} R + \Box^l R \Box^{n-l} R
\right) \nonumber \\
&- 2 \partial_0 \Box^l R \partial_0 \Box^{n-1-l} R\big)  =
\frac{-1}{8 \pi G}( G_{00} + \Lambda g_{00}).   \label{eq:3.4}
\end{eqnarray}

We are interested in  cosmological solutions for the universe with FLRW metric
and even in such simplified  case it is rather difficult to find solutions of the above equations.
To evaluate the above equations, the following Ans\"atze  were used:
\begin{itemize}
\item   Linear Ansatz: $\Box R = r R + s, $ where $r$ and $s$ are constants.
\item   Quadratic Ansatz:  $\Box R = q R^2, $ where $q$ is a constant.
\item   Qubic Ansatz:  $\Box R = q R^3, $ where $q$ is a constant.
\item    Ansatz $\Box^n R = c_n R^{n+1}, \,\, n\geq 1, $ where $c_n$ are constants.
\end{itemize}
In fact these Ans\"atze make some constraints on possible solutions, but on the other hand they
simplify formalism to find a particular solution.

\subsubsection{Linear Ansatz and Nonsingular Bounce Cosmological Solutions}

Using Ansatz  $\Box R = r R + s$ a few nonsingular bounce solutions for  the scale factor are found:
$a(t) = a_0 \cosh{\left(\sqrt\frac{\Lambda}{3}t\right)}$ (see \cite{biswas1,biswas2}),
 $\, a(t) = a_0 e^{\frac{1}{2}\sqrt{\frac{\Lambda}{3}}t^2}$ (see \cite{koshelev}) and $a(t) = a_0  (\sigma  e^{\lambda t} + \tau e^{-\lambda t} )$
 \cite{dragovich2}.
The first two consequences of this Ansatz are
\begin{equation}\label{nth degree}
\Box^{n} R = r^{n}(R +\frac sr ) , \, \, n\geq 1 , \,  \qquad \, \mathcal{F}(\Box)
R = \mathcal{F}(r) R + \frac sr(\FF(r)-f_0) ,
\end{equation}
which considerably simplify nonlocal term.

Now we can search for a solution of the scale factor $a(t)$ in the form of a linear combination
of $e^{\lambda t}$ and $e^{-\lambda t}$,  i.e.
\begin{equation} \label{sol:a}
a(t) = a_0  (\sigma  e^{\lambda t} + \tau e^{-\lambda t} ), \quad
0< a_0, \lambda,\sigma,\tau \in \Real .
\end{equation}
Then  the corresponding expressions for the Hubble parameter $H(t) = \frac{\dot{a}}{a},$ scalar curvature
$R(t) = \frac{6}{a^2} (a \ddot{a} + \dot{a}^2 + k) $ and $\Box R$ are:
\begin{equation} \begin{aligned} \label{sol:all}
H(t) &= \frac{\lambda  (\sigma  e^{\lambda t} - \tau e^{-
\lambda t}) } {\sigma  e^{\lambda t} + \tau e^{- \lambda t}}, \\
R(t) &= \frac{6 \left(2 a_0^2 \lambda ^2 \left(\sigma^2 e^{4 t
\lambda }+\tau ^2\right)+k e^{2 t \lambda }\right)}{a_0^2 \left(\sigma
e^{2 t \lambda }+\tau \right)^2},\\
\Box R &= -\frac{12 \lambda ^2 e^{2 t \lambda } \left(4 a_0^2
\lambda ^2 \sigma  \tau -k\right)}{a_0^2 \left(\sigma  e^{2 t \lambda
}+\tau \right)^2}.
\end{aligned} \end{equation}
We can rewrite $\Box R$ as
\begin{equation}\label{ansatz:1}
\Box R = 2\lambda^2 R - 24\lambda ^ 4 , \qquad r = 2\lambda^2 , \, \, s = - 24\lambda ^ 4.
\end{equation}

Substituting parameters $r$ and $s$  from \eqref{ansatz:1} into
\eqref{nth degree} one obtains

\begin{equation} \begin{aligned} \label{sol-all}
%\Box R &= 2\lambda^2 R - 24\lambda ^ 4 \\
\Box^n R &= (2\lambda^2)^n (R - 12\lambda ^2) , \, \, n \geq 1 ,  \\
\mathcal{F}(\Box) R &= \mathcal{F}(2 \lambda^2)R - 12
\lambda^2(\mathcal{F}(2 \lambda^2) - f_0).
\end{aligned} \end{equation}

Using this in \eqref{eq:3.3} and \eqref{eq:3.4} we obtain
\begin{align} \label{trace:11}
&36\lambda^2 \mathcal{F}(2 \lambda^2) (R - 12\lambda ^2) +
\mathcal{F}'(2 \lambda^2) \left( 4 \lambda^2 (R - 12\lambda ^2)^2
- \dot R^2 \right)  \nonumber \\
&-24 \lambda^2 f_0(R - 12 \lambda^2) = \frac{R - 4\Lambda}{8 \pi G} ,
\end{align}
\begin{align} \label{eom:2}
& (2 R_{00} + \frac{1}{2} R)\left( \mathcal{F}(2 \lambda^2)R - 12
\lambda^2(\mathcal{F}(2 \lambda^2) - f_0) \right)\nonumber \\ &-\frac{1}{2}\mathcal{F}' (2 \lambda^2) \left( \dot R^2 + 2 \lambda^2 (R - 12 \lambda^2)^2 \right)
-6\lambda^2(\FF(2\lambda^2)-f_0) (R-12\lambda^2)\nonumber \\ &+6 H \FF(2 \lambda^2) \dot R= - \frac{1}{8\pi G}( G_{00} - \Lambda) .
\end{align}

Substituting $a(t)$ from \eqref{sol:a} into equations \eqref{trace:11} and \eqref{eom:2} one obtains  two equations  as polynomials in $e^{2\lambda t}$.
Taking coefficients of these polynomials to be zero one obtains a system of equations and
their solution determines parameters  $a_0, \lambda,\sigma,\tau $ and yields some conditions for function $\mathcal{F}(2\lambda^2).$  For details
see \cite{dragovich2}.

\subsubsection{Quadratic  Ansatz and Power-Law Cosmological Solutions}

New  Ans\"atze $\Box R = r R, \,\, \Box R = q R^2$ and $\Box^n R = c_n R^{n+1},$ were introduced in \cite{dragovich1}
and they contain solution for $R =0$  which satisfies also equations of motion.   When  $k =0$ there is only static solution
$a= constant,$ and for $k=-1$  solution is $a(t) =  |t|.$

In particular, Ansatz $\Box R = q R^2$ is very interesting.
The corresponding differential equation for the Hubble parameter, if $k =0,$
is
\begin{equation}
\dddot{H} + 4\dot{H}^{2} + 7H \ddot{H} + 12 H^{2}\dot{H} + 6 q (
\dot{H}^2 + 4 H^2 \dot{H} + 4 H^4) = 0
\end{equation}
with solutions

\begin{equation} \label{Hn}
H_\eta(t) = \frac{2\eta+1}{3}\frac{1}{t + C_1}, \quad q_\eta
=\frac{6(\eta-1)}{(2\eta+1)(4\eta-1)},  \, \, \, \eta \in
\mathbb{R}
\end{equation}
and $H =\frac{1}{2}\frac{1}{t + C_1}$ with arbitrary coefficient
$q$, what is equivalent to the ansatz $\Box R = r R$ with $R = 0$.

The corresponding scalar curvature is given by
\begin{equation} \label{Rn}
R_\eta = \frac{2}{3}\frac{(2\eta+1)(4\eta-1)}{ (t+C_1)^{2}}, \, \,
\, \eta \in \mathbb{R}.
\end{equation}
By straightforward calculation one can show that $\Box^n R_n = 0$
when $n \in \mathbb{N}$. This simplifies the equations
considerably. For this particular case of solutions operator
$\mathcal{F}$ and trace equation \eqref{eq:3.3} effectively
become
\begin{align}
\label{operatorF:n} &\mathcal{F}(\Box) = \sum_{k=0}^{n-1} f_{k}\Box^k ,\\
 \label{trace:2}
 &\sum_{k=1}^{n+1} f_{k} \sum_{l=0}^{k-1} (\partial_{\mu}\Box^{l}R \partial^{\mu}\Box^{k-1-l}R + 2\Box^{l}R \Box^{k-l}R )
 + 6 \Box \mathcal{F}(\Box) R = \frac{R}{8 \pi G}.
\end{align}

In particular case $n=2$  the trace formula becomes
\begin{align} \nonumber
& \frac{36}{35}f_{0} R^{2} +  f_{1}(- \dot{R}^{2}+
\frac{12}{35}R^{3}) + f_{2}(-\frac{24}{35}R \dot{R}^{2} +
\frac{72}{1225}R^{4}) + f_{3} (-\frac{144}{1225}R^{2}\dot{R}^{2})\\
& = \frac{R}{8 \pi G}.  \label{trace:3}
\end{align}

Some details on all  the above three Ans\"atze can be found in \cite{dragovich1}.

\subsection{Nonlocal Model with Term $ R^{-1}
\mathcal{F}(\Box) R $}

This model was introduced recently \cite{dragovich3} and its action may be written in the form
\begin{equation} \label{eq-3.2.1}
S =  \int d^{4}x \sqrt{-g}\Big(\frac{R}{16 \pi G} +  R^{-1} \mathcal{F}(\Box) R  \Big),
\end{equation}
where $\mathcal{F}(\Box) = \sum_{n=0}^\infty f_n \Box^n $ and when $f_0 = -\frac{\Lambda}{8\pi G}$ it plays role of the
cosmological constant. For example, $\mathcal{F}(\Box)$ can be of the form $\mathcal{F}(\Box) = - \frac{\Lambda}{8\pi G} e^{-\beta \Box} .$

The nonlocal term $R^{-1} \mathcal{F}(\Box) R$ in \eqref{eq-3.2.1} is invariant under transformation $R \to C R.$ It means that effect of nonlocality does not
depend on the magnitude of scalar curvature $R,$ but on its spacetime dependence, and in the FLRW case is sensitive  only to dependence of $R$ on time $t$.
When $R= constant$ there is no effect of nonlocality, but only of $f_0$ what corresponds to cosmological constant.

 By variation of
action \eqref{eq-3.2.1} with respect to metric $g^{\mu\nu}$ one
obtains the equations of motion for $g_{\mu\nu}$
\begin{eqnarray}\label{eq-3.2.2}
&R_{\mu\nu} V - (\nmu\nnu - g_{\mu\nu} \Box)V -
\frac{1}{2} g_{\mu\nu} R^{-1} \mathcal{F}(\Box) R  \nonumber \\
&+ \sum_{n=1}^{\infty} \frac{f_n}{2} \sum_{l=0}^{n-1} \big(
g_{\mu\nu} \left( \partial_{\alpha} \Box^l(R^{-1})
\partial^{\alpha} \Box^{n-1-l} R + \Box^l(R^{-1}) \Box^{n-l} R
\right) \nonumber  \\
&- 2 \pmu \Box^l(R^{-1}) \pnu \Box^{n-1-l} R\big)   = - \frac{G_{\mu\nu}}{16 \pi G} ,  \label{eq-3.2.2} \\
&V = \FF(\Box) R^{-1} - R^{-2} \FF(\Box) R .  \nonumber
\end{eqnarray}
Note that operator $\Box$ acts not only on $R$ but also on $R^{-1}$. There are only two independent equations when metric is of the FLRW type.

The trace of the equation \eqref{eq-3.2.2} is
\begin{eqnarray}\label{eq-3.2.3}
&R V + 3 \Box V + \sum_{n=1}^{\infty} f_n \sum_{l=0}^{n-1} \left( \partial_{\alpha} \Box^l(R^{-1}) \partial^{\alpha} \Box^{n-1-l} R + 2 \Box^l(R^{-1}) \Box^{n-l} R \right) \nonumber \\
&-2 R^{-1} \FF(\Box) R  = \frac{R}{16 \pi G}.  \label{eq-3.2.3}
\end{eqnarray}

The $00$-component of \eqref{eq-3.2.2} is
\begin{eqnarray}
&R_{00} V - (\nabla_0\nabla_0 - g_{00} \Box)V -
\frac{1}{2} g_{00} R^{-1} \mathcal{F}(\Box) R \nonumber \\
&+ \sum_{n=1}^{\infty} \frac{f_n}{2} \sum_{l=0}^{n-1} \big(
g_{00} \left( \partial_{\alpha} \Box^l(R^{-1})
\partial^{\alpha} \Box^{n-1-l} R + \Box^l(R^{-1}) \Box^{n-l} R \right) \nonumber  \\
&- 2 \partial_0 \Box^l(R^{-1}) \partial_0 \Box^{n-1-l} R\big)  = - \frac{G_{00}}{16 \pi G}.  \label{eq-3.2.4}
\end{eqnarray}
These trace and $00$-component equations are equivalent for the FLRW universe in the equation of motion (20),
but they are more suitable for usage.

\subsubsection{Some Cosmological Solutions for Constant $R$}

We are interested in some exact nonsingular cosmological solutions for the scale factor $a(t)$ in \eqref{eq-3.2.2}.
The Ricci curvature $R$ in the above equations of motion   can be calculated by
expression $$R = 6 \left(\frac{\ddot{a}}{a} + \frac{\dot{a}^2}{a^2} + \frac{k}{a^2}\right).$$

\bigskip
\textbf{Case $k=0$,\,\, $a(t) = a_0 e^{\lambda t}$.}

We have $a(t) = a_0 e^{\lambda t}, \quad \dot{a} = \lambda a, \quad \ddot{a} = \lambda^2 a, \quad H = \frac{\dot{a}}{a} = \lambda$ and
$R = 6 \left(\frac{\ddot{a}}{a} + \frac{\dot{a}^2}{a^2} \right) = 12 \lambda^2.$  Putting  $a(t) = a_0 e^{\lambda t}$ in the above
equations \eqref{eq-3.2.3} and \eqref{eq-3.2.4}, they are satisfied with $\lambda = \pm \sqrt{\frac{\Lambda}{3}},$
where $\Lambda = - 8\pi G \, f_0 $ with  $f_0 < 0.$

\bigskip
\textbf{Case $k=+1$,\,\, $a(t) = \frac{1}{\lambda} \cosh{\lambda t}$.}

Starting with $a(t) = a_0 \cosh{\lambda t}, $ we have  $\dot{a} = \lambda a_0 \sinh{\lambda t}, \quad H = \frac{\dot{a}}{a} = \lambda \tanh{\lambda t}$
and $R = 6 \left(\frac{\ddot{a}}{a} + \frac{\dot{a}^2}{a^2} + \frac{1}{a^2}\right) = 12 \lambda^2$ if $a_0 = \frac{1}{\lambda} .$ Hence equation
 \eqref{eq-3.2.3} and \eqref{eq-3.2.4}
are satisfied for cosmic scale factor  $a(t) = \frac{1}{\lambda} \cosh{\lambda t} .$

\bigskip
In a similar way, one can obtain another solution:
\medskip

\textbf{Case $k=-1$,\, $a(t) = \frac{1}{\lambda} |\sinh{\lambda t}|$.}

\bigskip

Thus we have the following three cosmological solutions for $R = 12 \lambda^2$:
\begin{enumerate}
\item $k=0$, \,\, $a(t) = a_0 \, e^{\lambda t},$ nonsingular bounce solution.
\item $k=+1$,\,\,  $a(t) = \frac 1\lambda \, \cosh{\lambda t},$  nonsingular bounce solution.
\item $k=-1$, \,\, $a(t) = \frac 1\lambda \, |\sinh{\lambda t}|,$ singular cosmic solution.
\end{enumerate}
All of this solution have exponential behavior for large value of time $t$.

Note that in all the above three cases the following two tensors have also the same expressions:
\begin{equation}
 R_{\mu\nu} = \frac 14 R g_{\mu\nu}, \quad  \quad
G_{\mu\nu} = - \frac 14 R g_{\mu\nu}.
\end{equation}

Minkowski background space follows from the de Sitter solution  $k=0$, \, $a(t) = a_0 e^{\lambda t}.$ Namely, when $\lambda \to 0$ then
$a(t) \to a_0$ and $H =R=0.$

In all the above cases $\Box R = 0$ and thus coefficients $f_n ,\,\, n\geq 1$ may be arbitrary. As a consequence, in these cases nonlocality does not play a role.

\subsubsection{Some Power-Law Cosmological Solutions}

Power-law solutions in the form $a(t)= a_0 |t-t_0|^\alpha,$ have been investigated by some Ans\"atze in \cite{dragovich3} and without Ans\"atze \cite{dragovich4}.
The corresponding Ricci scalar and the Hubble parameter are:   $$R(t) = 6 \left(\frac{\ddot{a}}{a} + \frac{\dot{a}^2}{a^2} +
\frac{1}{a^2}\right) = 6\big(\alpha(2\alpha-1) (t-t_{0})^{-2} + \frac k{a_0^2}(t-t_{0})^{-2\alpha}\big)$$  $$H(t) = \frac{\dot{a}}{a} = \frac{\alpha}{|t-t_0|}.$$
Now $\Box = - \partial_t^2 - \frac{3\alpha}{|t - t_0|} \partial_t .$ An analysis has been performed for $\alpha \neq 0, \, \frac{1}{2},$ and also
$ \,  \alpha \to 0, \quad \alpha \to \frac{1}{2}$ for $ \, k=+1, -1, 0.$  For details, the reader refers to \cite{dragovich3,dragovich4}.

\section{Discussion and Concluding Remarks}

To illustrate the form of the abave nonlocality \eqref{eq:3.1} it is worth to start from exact effective Lagrangian at the tree level for
$p$-adic closed and open scalar strings.
This Lagrangian is as follows (see, e.g. \cite{freund}):
\begin{align}
L_p = &- \frac{m^D}{2g^2}\frac{p^2}{p-1} \varphi p^{-\frac{\Box}{2m^2}} \varphi - \frac{m^D}{2h^2}\frac{p^4}{p^2-1} \phi p^{-\frac{\Box}{4m^2}} \phi
 +\frac{m^D}{h^2}\frac{p^4}{p^4-1} \phi^{p^2 +1}  \nonumber \\ & - \frac{m^D}{g^2}\frac{p^2}{p^2-1} \phi^{\frac{p(p-1)}{2}} +
 \frac{m^D}{g^2}\frac{p^2}{p^2-1} \varphi^{p+1} \phi^{\frac{p(p-1)}{2}}, \label{0.1}
\end{align}
where $\varphi$ denotes open strings, $D$ is spacetime dimensionality (in the sequel we shall take $D=4$), and $g$ and $h$ are coupling constants
for open and closed strings, respectively.
Scalar field $\phi(x)$ corresponds to closed $p$-adic strings and could be related to gravity scalar curvature as $\phi =f(R)$, where $f$ is an
appropriate function. The corresponding equations of motion are:
\begin{align}
p^{-\frac{\Box}{2m^2}} \varphi = \varphi^{p} \phi^{\frac{p(p-1)}{2}} , \quad   p^{-\frac{\Box}{4m^2}} \phi =  \phi^{p^2} + \frac{h^2}{2 g^2} \frac{p-1}{p}
\phi^{\frac{p(p-1)}{2}-1} \left( \varphi^{p+1} - 1 \right). \label{0.2}
\end{align}
There are the following constant vacuum solutions: $(i)\, \varphi = \phi = 0 $, $(ii)\, \varphi = \phi = 1 $ and $(iii)\, \varphi = \phi^{-\frac{p}{2}}
= constant.$

In the case that the open string field $\varphi = 0,$ one obtains equation of motion only for closed string $\phi .$  One can now
construct a toy nonlocal gravity model supposing that closed scalar string is related to the Ricci scalar curvature as
$\phi = - \frac{1}{m^2} R = - \frac{4}{3 g^2} (16\pi G) R. $
Taking  $p = 2,$ we obtain the following Lagrangian for gravity sector:
\begin{align}
\mathcal{L}_g  =   \frac{1}{16\pi G}\, R  - \frac{8}{3} \frac{C^2}{h^2} R\, e^{-\frac{\ln 2\, \Box}{4m^2}}\, R
 -\frac{1024}{405 g^6 h^2} (16\pi G)^3 R^{5}  . \label{0.3}
\end{align}
 To compare third term  to the first one in \eqref{0.3}, let us note that
 $(16\pi G)^3 R^{5} =  (16\pi G R)^4 \frac{R}{16\pi G}. $ It follows that $(G R)^4$ has to be dimensionless after rewriting it using constants $c$ and $\hbar.$
As Ricci scalar $R$ has dimension $Time^{-2}$ it means that $G$ has to be replaced by the Planck time as $t_P^2 = \frac{\hbar G}{c^5} \sim 10^{-88} s^2. $ Hence
$(G R)^4 \to (\frac{\hbar G}{c^5} R)^4 \sim 10^{-352} R^4 $ and third term in \eqref{0.3} can be neglected with respect to the first one, except when $R \sim t_P^{-2}.$ The nonlocal model with only first two terms corresponds to case considered above in this article. We shall consider this model including $R^5$ term elsewhere.

It is worth noting that the above two models with nonlocal terms $R \mathcal{F}(\Box) R$ and $R^{-1} \mathcal{F}(\Box) R$ are equivalent in the case when $R = constant,$
because their equations of motion have the same solutions. These solutions do not depend on $\mathcal{F}(\Box) - f_0 .$  It would be useful to find
cosmological solutions which have definite connection with the explicit form of nonlocal operator $\mathcal{F}(\Box).$

Let us mention that many properties of \eqref{eq:3.1} and its extended quadratic versions have been considered, see
\cite{biswas3,biswas4,biswas3+,koshelev1,koshelev2}.

Nonlocal model \eqref{eq-3.2.1} is a new one and was not considered before \cite{dragovich3}, it seems to be important and deserves further investigation. There are
some gravity models modified by term $R^{-1},$ but they are neither nonlocal nor pass Solar system tests, see e.g. \cite{kamionkowski}.

Note that nonlocal cosmology is related also to cosmological models in which matter sector contains nonlocality (see, e.g.
 \cite{arefeva0,arefeva,calcagni1,barnaby,koshelev-v,arefeva-volovich,dragovich,dragovich-d}).
String field theory and $p$-adic string theory models have played significant role in motivation and construction of such models.

Nonsingular bounce cosmological solutions are very important (as reviews on bouncing cosmology, see e.g.  \cite{novello,brandenberger})
and their progress in nonlocal gravity may be a further step towards cosmology of the cyclic universe \cite{steinhardt}.

\begin{acknowledgement}
Work on this paper was supported by Ministry of Education, Science and Technological Development of the Republic of Serbia, grant No 174012.
The author thanks Prof. Vladimir Dobrev for invitation to participate and give a talk, as well as for hospitality, at the X International Workshop ``Lie Theory
and its Applications in Physics'', 17--23 June 2013, Varna, Bulgaria. The author also thanks organizers of the Balkan Workshop BW2013 "Beyond Standard Models" (25-29.04.2013, Vrnja\v cka Banja, Serbia), Six Petrov International Symposium on High Energy Physics, Cosmology
and Gravity (5-8.09.2013, Kiev, Ukraine) and Physics Conference TIM2013 (21-23.11.2013, Timisoara, Romania), where some results on modified gravity and its cosmological solutions were presented. Many thanks also to my collaborators Zoran Rakic, Jelena Grujic and Ivan Dimitrijevic,
as well as to Alexey Koshelev and Sergey Vernov for useful discussions.
\end{acknowledgement}

\end{document}